\documentclass[article,showpacs,showkeys,superscriptaddress]{revtex4}

\usepackage{amsmath}
\usepackage{bbold}
\usepackage{amsfonts}
\usepackage{amssymb}
\usepackage{pbsi}
\usepackage[T1]{fontenc}
\usepackage{hyperref}
\usepackage{xcolor}
\usepackage{graphicx}
\usepackage{subfig}
\usepackage{float}

\usepackage{color}

\def \ee{\end{equation}}
\def \be{\begin{equation}}
\def \eea{\end{eqnarray}}
\def \bea{\begin{eqnarray}}

\begin{document}

\title{Extended  
BMT equations
and the anomalous magnetic moment}

\author{Felipe A. Asenjo}
\email{felipe.asenjo@uai.cl (corresponding author)}
\affiliation{Facultad de Ingenier\'ia y Ciencias,
Universidad Adolfo Ib\'a\~nez, Santiago 7491169, Chile.}
\author{Benjamin Koch}
\email{benjamin.koch@tuwien.ac.at}
\affiliation{Institut f\"ur Theoretische Physik,
 Technische Universit\"at Wien,
 Wiedner Hauptstrasse 8--10,
 A-1040 Vienna, Austria}
 \affiliation{Atominstitut, Technische Universit\"at Wien,  Stadionallee 2, A-1020 Vienna, Austria}
 \affiliation{ Instituto de F\'isica, Pontificia Universidad Cat\'olica de Chile, 
Casilla 306, Santiago, Chile}

\begin{abstract}
We propose a generalized form of the Thomas-Bargmann-Michel-Telegdi equations. These are first-order in both electric and magnetic fields and retain the conventional conserved quantities and constraints. Within this novel phenomenological framework, we delve into an archetypal measurement scenario. Specifically, we scrutinize the contributions to the standard definition of the anomalous magnetic moment observable, highlighting each new correction term.
\end{abstract}


\maketitle

\tableofcontents
\section{Introduction}

\subsection{Motivation}

At the fundamental level, all known non-Gravitational interactions are described in terms of quantum field theory.
Nevertheless, in many cases it is sufficient and useful to work with particles which follow a classical equation of motion, defined in a classical  background.
The best known example for this is the Lorentz force, which describes the motion of a charged structureless test charge in an electromagnetic background.
In nature, there is, however, no particle-like object which only carries charge and momentum. Fundamental particles like electrons or muons carry an internal angular momentum, which also couples to the external fields. The classical motion of spinning particles is described by
Thomas-Bargmann-Michel-Telegdi (BMT) equations~\cite{Frenkel:1926zz,Thomas:1926dy,Bargmann:1959gz}.
The study of the coupling that appears in these equations has attracted the increasing attention of physicists for almost a century. While in the case of fundamental standard-model particles, there is good reason to believe that the BMT equation and the associated couplings are very good approximative descriptions, the situation changes for composite particles like protons, neutrons, or other hadronic states. The complicated internal structure of these particle states,  allows for the possibility of additional couplings, such as multipole moments. Thus, 
one way of learning the internal structure of composite particles or to discover novel properties of fundamental particles is to explore these generalizations of the BMT equations.

In this paper we explore novel self-consistent generalizations of the BMT equations. Thus we provide a tool that is potentially useful for the study of composite and fundamental particles. As exemplary observable we use the anomalous magnetic moment to analyze our generalized BMT equations. 

The text is structured as follows.
In the following subsection \ref{ss_BMT} we revisit the five conditions that lead to the BMT equations and show which observables allow to determine the anomalous magnetic moment.
In subsection \ref{bmtpluselectridopole} we revisit the literature on generalizations of the BMT equations. In the last subsection \ref{ss_idea} of the introduction we explain the idea behind our approach.
In section \ref{se_BMTpp} we obtain our generalized BMT equations in three steps by postulating an ansatz \ref{ss_raw}, imposing constraints \ref{ss_conserved}, and deriving the final set of equations \ref{generalfinalequations}.
We explore these equations,
by using exact analytic methods for special scenarios \ref{sec_exact} and by using perturbation theory \ref{sec_perturbation}.

\subsection{BMT equations and $g-2$}
\label{ss_BMT}

The classical motion of particles with spin in
external electromagnetic fields is described by the Thomas-Bargmann-Michel-Telegdi (BMT) equations~\cite{Frenkel:1926zz,Thomas:1926dy,Bargmann:1959gz}.
Even now, almost a century after their foundations, these equations are extensively used, e.g. to describe the motion of spinning particles in storage rings and to extract the corresponding experimental information, such as the anomalous magnetic moment of the muon~\cite{Miller:2007kk,Miller:2012opa,Muong-2:2021ojo,Muong-2:2023ojo}
\be\nonumber
a_\mu = \frac{g-2}{2}.
\ee
From the phenomenological perspective, the equations are not derived from the Standard Model of particles, nevertheless they are used to test the parameters of this model. 
The free parameters
of the BMT equations,  that have to be determined by experiment, are $(S^2,\, P^2,\, c,\, e,\, g)$, corresponding to the square of the magnitude of the spin, the square of mass, velocity of light, charge, and magnetic moment. 
 The derivation of the BMT equations relies in the following conditions:
\begin{enumerate}
    \item The dynamical quantities for the particle are the momentum $P^\mu(\tau)$ and spin vector $S^\mu(\tau)$.
   \item 
    The covariant equations are first order in the electromagnetic field strengths.
    \item Spin, and momentum are orthogonal to each other and conserved.
    \item Both equations are independent of the dual field strength tensor $F^{*\mu \nu}$ (unless it can be reduced to $F^{\mu \nu}$ without an epsilon tensor)
    \item 
    The dynamical equations for $P^\mu$ are independent of $S^{\mu}$.
    \item 
    The field equations do not depend on derivatives of the field strength.
\end{enumerate}

These conditions imply for example that the spin equations reduce to the Thomas precession in the rest frame.
Note that these equations can alternatively be derived as equations of motion of a classical spherical top~\cite{Hanson:1974qy}.
The resulting BMT equations~\cite{Frenkel:1926zz,Thomas:1926dy,Bargmann:1959gz} read
\be\label{eq_BMT1}
m\frac{d p^\mu}{d \tau} =
\frac{e}{c} F^{\mu \nu}p_\nu
\ee
and
\bea\label{eq_BMT2}
m\frac{d S^\mu}{d \tau}&=&
\frac{ge}{2c} F^{\mu \nu}S_\nu 
+e\frac{(g-2)}{2 c m^2} p^\mu (FSp).
\eea
where the shorthand notation $(FSp)=-(FpS)=F^{\mu \nu}S_\mu p_\nu$ was used. These equations imply that if initially $S^2=S^\mu S_\mu$,
$m^2=p^\mu p_\mu$, and $S^\mu p_\mu=0$,
then these three relations are 
constants of motion.
One can define the 4-velocity $u^\mu=p^\mu/m$, the laboratory velocity $v^\mu=u^\mu \frac{d\tau}{dt}$,
and the
local spin pseudo-vector
\be\label{eq_zeta}
\vec \zeta= \vec S- \frac{\gamma}{ (1+\gamma)}
{\vec v (\vec v \cdot \vec S)},
\ee
where $\gamma^2=({1-\vec v^2/c^2})^{-1/2}$.

It is then found that in a constant magnetic field (with no electric field), the first equation~(\ref{eq_BMT1}) reads
\be
m\frac{d\vec v}{dt}= \frac{e}{\gamma} \vec v \times \vec B\, ,
\ee
while the second equation (\ref{eq_BMT2}) can, under the same conditions, be written as
\be
\frac{d \vec \zeta}{dt}=\frac{ge}{2 m c \gamma} \vec \zeta \times \vec B- (g-2)
\frac{e}{2 m c}\frac{\gamma}{\gamma +1}
\vec \zeta \times \left[\frac{\vec v}{c}\times\left(\frac{\vec v}{c} \times \vec B\right)\right].
\ee
If $\vec v \cdot \vec B=0$, we find that the two oscillation frequency vectors are parallel to $\vec B$ such that 
\begin{equation}
 \frac{d \vec v}{dt}=\vec v \times \vec\omega_p   \, ,\qquad 
 \frac{d \vec \zeta}{dt}=\vec\zeta\times \vec\omega_\zeta   ,
\end{equation}
with
\bea
\vec \omega_p&=& \frac{e}{m c \gamma} \vec B, \label{eq7fre}\\
\vec \omega_\zeta&=& \frac{ge}{2 m c \gamma} \vec B +  (g-2)
\frac{e}{2 m c}\frac{\gamma}{\gamma +1} \frac{\vec v^2}{c^2} \vec B.\label{eq8fre}
\eea
Thus, 
the difference between these two characteristic frequency vectors, 
projected onto the $\vec B$ direction,
gives the anomalous magnetic moment
\be\label{eq_amu1}
a_\mu= \frac{g-2}{2}=\frac{m c}{e\vec B^2 }
(\vec \omega_\zeta - \vec \omega_p)\cdot \vec B .
\ee

\subsection{BMT$^+:$ Existing generalization}
\label{bmtpluselectridopole}

Over the years, the BMT equations have been generalized to include electric dipole moment and contributions from parity violation,  (BMT$^+$)~\cite{Nelson:1959zz,Khriplovich:1998zq,Fukuyama:2013ioa,Porshnev:2021xdz,Nikolaev:2020wsj}. 
 The generalized spin equation reads
\bea\label{eq_BMT2EDM}
m\frac{d S^\mu}{d \tau}=
\frac{ge}{2c} F^{\mu \nu}S_\nu 
+e\frac{(g-2)}{2 c m^2} p^\mu (FSp)  - 2d m\left(F^{*\mu \nu} S_\nu - u^\mu 
F^{*\alpha \beta}u_\alpha S_{\beta}\right),
\eea
where $F^{* \mu \nu}$ is the dual field strength tensor.
This generalization contains the electric dipole moment $d$, as additional free parameter~\cite{Anastassopoulos:2015ura,Chupp:2017rkp,Dekens:2018bci}. 
This parameter can be expected to arise for
composite particles.
Gravity coupling has also been extensively discussed~\cite{Vergeles:2022mqu,Silenko:2018zct}, but shall not be considered in the following. 
In the model described by Eq.~(\ref{eq_BMT2EDM}), the momentum equation remains unaltered. It further fulfills all the proper constraints of the theory.
Interestingly, the  electric dipole moment parameter appears in 
the difference of eigenfrequencies
\be\label{eq_gm2usual}
\vec \omega_\zeta - \vec \omega_p= 
\frac{e}{mc} \left[\frac{g-2}{2} \vec B+ \frac{\eta}{2}\left(\frac{\vec v}{c} \times \vec B\right)\right],
\ee
where $d={e \eta}/({4m})$.
Even though there is a correction in the above difference, 
it would not make a difference in the definition (\ref{eq_amu1}).
Further generalizations include  the extensively studied gradients of external fields~\cite{Metodiev:2015gda},
 self-force 
corrections to the Lorentz force~\cite{Abraham:1903,Lorentz:1915,Dirac:1938nz}, or bremsstrahlungs corrections~\cite{Pardy:2008di}.
An Amperian (current loop) and Gilbertian
(magnetic monopole) have been considered in~\cite{Rafelski:2017hce,Formanek:2020ojr,Steinmetz:2023ucp}. These works, also pesent an elegant way to introduce field gradients by introducing the covariant magnetic potential
\be
B_\mu \equiv F_{\mu \alpha}^* S^\alpha,
\ee
which by construction satisfies 
\be
\partial_\mu B^\mu=0.
\ee
For this vector, one then defines a covariant ``field strength'' via
\be
G^{\alpha \beta}=\partial^\alpha B^\beta-\partial^\beta B^\alpha.
\ee

\subsection{The idea}
\label{ss_idea}

The aim of this paper to is 
revisit the six conditions that lead to the BMT equations. In particular, we are interested in scenarios where one can relax some of these conditions and still obtain meainingful generalizations of the BMT equations. 
Some of the conditions rest on important physical principles, while others serve for mathematical ``simplicity''. Thus, some are more suitable for being relaxed than others:
\begin{itemize}
    \item Condition 1 defines the dynamical variables, thus we want to keep it.
    \item Condition 2 contains two parts. First, there is the covariance. We want to keep covariance since it linked to Lorentz symmetry, which is essential for all established fundamental descriptions of nature. Second, demanding first order in electromagnetic field strenght is not strictly necessary, but it is a very good approximation to most experimental scenarios where higher order field strength is negligible. Thus, we keep this condition, but we note that the second part of this condition can eventually also be relaxed.
    \item Condition 3 arises from symmetries and their associated conservation laws. For example it emerges naturally as consequence of a fundamental theory for classical spinning massive particles in flat spacetime \cite{Hanson:1974qy}. Thus, we opt to avoid anomalies and keep this condition.
    \item Condition 4 is neither protected by a fundamental symmetry, nor is it necessary for the definition of the dynamical variables. We will thus try to relax this condition.
    \item Condition 5 serves first and foremost to simplify the resulting system of differential equations. We take the risk of getting more complicated equations and relax this condition.
    \item Condition 6 also only serves the mathematical simplicity. In fact this condition has been dropped in known applications. Thus, we will also drop this condition.
\end{itemize}

Thus, we will relax conditions 4-6 and derive consistent dynamical equations from the conditions 1-3.
Since the resulting equations turn out to be a further generalization of the BMT$^+$ equations, we shall refer to them as BMT$^{++}$ equations.
The finding  will then lead also to a generalization of Eq.~(\ref{eq_BMT2EDM}), and the corresponding equations for the momentum.
Our results are not directly derived from a the
standard model of particle physics.
Probably, 
most of our couplings are either zero or loop suppressed when contrasted with fundamental particles of the standard model, since fundamental particles neither do  possess exotic multipole moments, nor do they have internal structure or susceptibility.
Nevertheless, composite particles can have all sorts of internal structure and multipole moments which in principle can be mapped to our couplings. 
Thus, our equations are more likely to be relevant for the analysis of composite particles with non-trivial internal structure. 
To exemplify how the novel couplings interfere with a correct interpretation of experimental results we revisit an idealized measurement of the anomalous magnetic moment.

We want to add a word of caution, before diving into the action. Even though we use the anomalous magnetic moment $g$, to analyze the dynamics of our equations, we do not claim
that these additional couplings are present for the muon.

\section{BMT$^{++}$: Generalized equations from scratch} 
\label{se_BMTpp}

We will now propose a generic set of dynamical equations for $S^\mu$ and $p^\mu$ and 
then constrain the couplings of these equations such that 
the expected conservation laws still hold.

\subsection{Raw set of equations}
\label{ss_raw}

Let's consider the most general Lorentz invariant equations for $S^\mu$ and $p^\mu$, which fulfill conditions 1 and 2:
\bea
m\frac{d p^\mu}{d \tau} &=&
\alpha F^{\mu \nu}p_\nu + \alpha^* F^{*\mu \nu}p_\nu
+
\beta F^{\mu \nu}S_\nu + \beta^* F^{*\mu \nu}S_\nu
\\ \nonumber &&
+
\kappa\, p^\mu (FSp) + \kappa^*\,  p^\mu (F^*Sp)
+\delta S^\mu (FSp) + \delta^* S^\mu (F^*Sp)
\\ \nonumber &&
+
\tilde \alpha G^{\mu \nu}p_\nu + \tilde \alpha^* G^{*\mu \nu}p_\nu
+
\tilde \beta G^{\mu \nu}S_\nu + \tilde \beta^* G^{*\mu \nu}S_\nu
\\ \nonumber &&
+
\tilde \kappa p^\mu (GSp) + \tilde \kappa^* p^\mu (G^*Sp)
+\tilde \delta S^\mu (GSp) + \tilde\delta^* S^\mu (G^*Sp)
\, ,
\eea
where each term has a free real parameter as prefactor. Here $F^*Sp=F^{*\mu \nu}S_\mu p_\nu$, and similar expressions for relations involving tensor $G^{\mu\nu}$. 
 In the above (and in the following) equation, the superscript $^*$ in the 
 real parameters are used to denote their relation to terms involving the
dual field strength tensor.
Analogously, the equation for the spin reads
\bea
m\frac{d S^\mu}{d \tau}&=&
A F^{\mu \nu}S_\nu + A^*F^{*\mu \nu}S_\nu 
+C F^{\mu \nu}p_\nu + C^*F^{*\mu \nu}p_\nu 
\\ \nonumber &&
+B p^\mu (FSp)+ B^* p^\mu (F^*Sp)
+D S^\mu (FSp)+ D^* S^\mu (F^*Sp)
\\ \nonumber &&
+\tilde A G^{\mu \nu}S_\nu + \tilde A^*G^{*\mu \nu}S_\nu 
+\tilde C G^{\mu \nu}p_\nu + \tilde C^*G^{*\mu \nu}p_\nu
\\ \nonumber &&
+\tilde B p^\mu (GSp)+ 
\tilde B^* p^\mu (G^*Sp)
+\tilde D S^\mu (GSp)+ \tilde D^* S^\mu (G^*Sp).
\eea

Thus, there are in total 32 parameters.
One can redefine $p^\mu$ and 
$S^\mu$ in arbitrary linear combinations. Note that we do not assume that $p^\mu$ is a velocity but rather a momentum and we further do not assume that one can assign simple parity symmetry to neither the momentum, nor spin vector. 
Making use of this possibility to mix $S^\mu$ and $p^\mu$ one can fix two prefactors. We choose
\bea\label{eq_choice0}
C&=&\tilde C=0\\ \nonumber
\beta &=&\tilde \beta=0,
\eea
because these allow to make a distinction between $S^\mu$ and $p^\mu$ in the absence of parity symmetry.
Following this selection, we are left with 28 remaining parameters.

\subsection{Conserved quantities}
\label{ss_conserved}

We would like to have the following 
conserved quantities
\bea\label{eq_conserved}
p_\mu p^\mu &=& P^2=m^2\\ \nonumber
S_\mu S^\mu &=& S^2\\ \nonumber
S_\mu p^\mu &=& 0,
\eea
where $S^2<0$.  These equations are the mathematical consequence of condition 
3.
These conservation laws can be guaranteed by using contractions of the equations of motion
\bea
p_\mu \frac{d p^\mu}{d \tau}&=&0,\\
S_\mu \frac{d S^\mu}{d \tau}&=&0,\\
p_\mu \frac{d S^\mu}{d \tau}+S_\mu \frac{d p^\mu}{d \tau}&=&0.
\eea
Since electric and magnetic fields are independent, each of these relations gives two conditions on the parameters.
Since further all terms with $F^{\mu \nu}$ are doubled with the corresponding terms containing field gradients $G^{\mu \nu}$,
all relations that must for coefficients without tilde also must hold for the corresponding coefficients with tilde
\bea\label{eq_conditions}
\kappa=\tilde \kappa &=&0 ,\\
\kappa^* &=& \frac{ \beta^*}{P^2},\\
\tilde\kappa^* &=& \frac{ \tilde\beta^*}{P^2},\\
B&=&\frac{A -\alpha - S^2 \delta }{P^2},\\
\tilde B&=&\frac{\tilde A -\tilde\alpha - S^2\tilde \delta }{P^2},\\
B^*&=&\frac{A^* -\alpha^* - S^2 \delta^* }{P^2},\\
\tilde B^*&=&\frac{\tilde A^* -\tilde \alpha^* - S^2 \tilde \delta^* }{P^2},\\
C^*&=&-D^*S^2\\
\tilde C^*&=&-\tilde D^*S^2,,\\
D=\tilde D&=&0.
\eea
Using these relations, there are 16 parameters left.
It is convenient to use the usual definitions of the electric coupling in terms of $\alpha$, the magnetic dipole moment and the magnetic moment in terms of $A$
and the analogous relations for the parameters of the field gradients
 \bea
\alpha &=& e/c\, ,\\
\tilde \alpha &=& \tilde e/c\, ,\\
A&=& \frac{g e}{2c}\,,\\
\tilde A&=& \frac{\tilde g \tilde e}{2c}\, .
\eea
Taking these as part of the usual BMT equations,
there remain 12 additional parameters, which can lead to deviations from the usual BMT equations.
This is, of course a mess.
Thus, it is useful to study different sub-cases as benchmarks.

\subsection{Final equations}
\label{generalfinalequations}

The full set of equations that meets our conditions is thus
\bea\label{eq_eom2U}
m\frac{d p^\mu}{d \tau} &=&
\frac{e}{c} F^{\mu \nu}p_\nu+\alpha^* F^{*\mu \nu}p_\nu
 + \beta^* F^{*\mu \nu}S_\nu
 + \frac{\beta^*}{P^2} p^\mu (F^*Sp)
+\delta S^\mu (FSp) + \delta^* S^\mu (F^*Sp)\\ \nonumber
&&+
\frac{\tilde e}{c} G^{\mu \nu}p_\nu+\tilde \alpha^* G^{*\mu \nu}p_\nu
 + \tilde \beta^* G^{*\mu \nu}S_\nu
 + \frac{\tilde \beta^*}{P^2} p^\mu (G^*Sp)
+\tilde \delta S^\mu (GSp) + \tilde \delta^* S^\mu (\tilde G^*Sp)\, ,
\eea
and
\bea\label{eq_eom2S}
m\frac{d S^\mu}{d \tau}&=&
\frac{ge}{2c} F^{\mu \nu}S_\nu + A^*F^{*\mu \nu}S_\nu 
- D^* S^2 F^{*\mu \nu}p_\nu 
\\ \nonumber &&
+e\frac{(g-2)-2 \delta  c S^2/e }{2 c P^2} p^\mu (FSp)+ \frac{A^*-\alpha^*-S^2 \delta^*}{P^2} p^\mu (F^*Sp)
+ D^* S^\mu (F^*Sp)\\ \nonumber
&&+
\frac{\tilde g\tilde e}{2c} G^{\mu \nu}S_\nu + \tilde A^*G^{*\mu \nu}S_\nu 
-\tilde  D^* S^2 F^{*\mu \nu}p_\nu 
\\ \nonumber &&
+\tilde e\frac{(\tilde g-2)-2 \tilde \delta  c S^2/ \tilde e }{2 c P^2} p^\mu (GSp)+ \frac{\tilde A^*-\tilde \alpha^*-S^2 \tilde \delta^*}{P^2} p^\mu (G^*Sp)
+\tilde  D^* S^\mu (G^*Sp),
\eea
combined with the conserved relations (\ref{eq_conserved}).

In what follows, we will study these equations for 
small field gradients
\be
G^{\mu \nu}\approx 0,
\ee
or equivalently, we will set all tilde couplings to zero.
This does not mean that field gradients are unimportant in practize, 
it only means that we want to show that the new couplings can lead to interesting corrections to the BMT equations, even in the absence of field gradients. This conclusion will of course not be altered, when field gradients are allowed.

Note that in the usual extended equations (\ref{eq_BMT2EDM}),
the electric dipole moment is a common factor of two different terms, while in (\ref{eq_eom2S}) the two terms have different prefactors, depending on $(A^*, \; \alpha^*, \; \delta^*,\; S^2,\; P^2)$.
Only for $\alpha^*=\delta^*$
the two prefactors agree and one can identifa the familiar electric dipole moment as $d=-A^*/2$.
The equations (\ref{eq_eom2U} and \ref{eq_eom2S}) are the main result of our paper.
They contain 8 free coupling parameters,
which consist in the 2 familiar couplings
 $\left\{e,\; g\right\}$ and the 6 additional couplings
 $\left\{ \alpha^*,\; \beta^*, \;\delta, \; \delta^*,\; A^*,\; D^*\right\}$.
Some of these extensions have been discussed in the literature, but 
to our knowledge, this is the most exhaustive generalization of the BMT equations so far.
The 8 independent couplings are, in theory, all experimentally measurable. This means they can either have constraints placed on the corrections or have their values ascertained. To facilitate such a phenomenological exploration, it's essential to extract predictions from Eqs.~(\ref{eq_eom2U}) and (\ref{eq_eom2S}). These predictions can then be juxtaposed with experimental observations.

In a direct approach, the goal is to solve these equations as comprehensively as possible. However, these intertwined differential equations prove to be exceptionally challenging to solve, even when considering rudimentary electromagnetic field configurations. A predominant challenge arises from the inherent non-linearity of these equations. To navigate these complexities and derive practical solutions, we employ the following strategies:
\begin{itemize}
    \item Derive exact analytical solutions
    for a reduced subset of couplings. An example for this strategy will be given in the next section.
    \item  Use known solutions for a subset of the couplings  and a perturbative approach to reduce the equations to linear order. A simple example of this approach will be presented in the subsequent section.
    \item Use numerical integration. This approach is left for future studies.
\end{itemize}

\section{Polishing I, exact results}
\label{sec_exact}

First, we want to gain exact analytic insight on the equations (\ref{eq_eom2U}, \ref{eq_eom2S}) by studying 
particularly interesting special cases.

\subsection{Anomalous magnetic moment}

The BMT equations are recovered from 
(\ref{eq_eom2U}, \ref{eq_eom2S}), for the case ($A^*=D^*=\alpha^*=\beta^*=\delta^*=\delta=0$).
One of the most well-known exact results for these equations is 
given for constant electric and magnetic fields ($\vec E,\; \vec B$).
In this case, it turns out that the 
the momentum vector $\vec P(\tau)$ and the spin pseudo-vector $\vec S$  have  characteristic frequencies given by Eqs.~\eqref{eq7fre} and \eqref{eq8fre}.


\subsection{Electric dipole moment and magnetic monopole}

A relativistic particle with spin and electric dipole moment  \cite{Nelson:1959zz,Khriplovich:1998zq,Fukuyama:2013ioa,Porshnev:2021xdz,Nikolaev:2020wsj} was discussed in Sec.~\ref{bmtpluselectridopole}. In the current general description of Sec.~\ref{generalfinalequations}, the electric dipole moment can be included in
 the special case when 
 $\alpha^*=\beta^*=\delta=\delta^*=D^*=0$ in Eqs.~(\ref{eq_eom2U})
and (\ref{eq_eom2S}).

This result is exact. 
However, a more general exact result can be obtained when only
   the parameters $\beta^*=\delta=\delta^*=D^*=0$. This corresponds to a system including the parameter set ($e,g,\alpha^*, A^*$).
In this case, Eqs.~(\ref{eq_eom2U})
and (\ref{eq_eom2S}) can be re-written as
\bea\label{eq_BMTgen1}
m\frac{d p^\mu}{d \tau} &=&\frac{e}{c} Z_2^{\mu \nu}p_\nu
\eea

and

\bea\label{eq_BMTgen2}
m\frac{d S^\mu}{d \tau}&=&
\frac{ge}{2c} Z_1^{\mu \nu}S_\nu +\frac{e}{2 c m^2} p^\mu \left[\left(g Z_1-2 Z_2\right)Sp\right],
\eea
where
\begin{eqnarray}
    Z_1^{\mu\nu}&=&F^{\mu\nu}+\frac{2 c}{g e}A^* F^{*\mu\nu}\, ,\nonumber\\
    Z_2^{\mu\nu}&=&F^{\mu\nu}+\frac{c}{e}\alpha^* F^{*\mu\nu}\, .
\end{eqnarray}

 From the above definition for $Z_2^{\mu\nu}$, we can see  that $\alpha^*$ is  related to a magnetic monopole property of a particle, which produces a force proportional to $F^{*\mu\nu}p_\nu$. Similarly, from $Z_1^{\mu\nu}$, we have that the parameter $A^*$ is related to the electric dipole moment of a particle, as it allows a spin precession proportional to $F^{*\mu\nu}S_\nu$.

Eqs.~\eqref{eq_BMTgen1} and \eqref{eq_BMTgen2} are remarkably similar to BMT equations \eqref{eq_BMT1} and \eqref{eq_BMT2}. Thereby, we can readily obtain that, for non-vanishing velocity, the difference between characteristic frequencies is
\be\label{eq_aMonopole}
\vec \omega_\zeta - \vec \omega_p= 
\frac{e}{mc} \left[\frac{g}{2} \vec B_1-\vec B_2-\frac{\vec v}{c}\times \left(\frac{g}{2}\vec E_1-\frac{\gamma^2}{\gamma^2-1}\vec E_2\right)-\frac{\gamma}{\gamma+1}\frac{\vec v}{c}\left(\frac{g}{2}\vec B_1-\vec B_2\right)\cdot\frac{\vec v}{c}\right],
\ee
where $\vec B_i$ ($\vec E_i$) is the magnetic (electric) part of the tensor $Z_i^{\mu\nu}$ (with $i=1,2$). Notice that when $\vec B_1=\vec B_2$ (and $\vec E_1=\vec E_2$) we recover the result from BMT theory.

Under the previous condition of a constant external magnetic field (with no electric field), then $\vec B_1=\vec B_2=\vec B$, $\vec E_1=-({\eta}/{g}) \vec B$, and $\vec E_2=-\chi \vec B$,
where we have chosen  
$A^*=e\eta/(2 c)$ (with the electric dipole moment coefficient $\eta$ defined in Sec.~\ref{bmtpluselectridopole}), and $\alpha^*=e\chi/c$, where $\chi$ is the effective coupling parameter for magnetic monopoles.
In this case, we find that
\begin{equation}
    \vec \omega_\zeta - \vec \omega_p=\frac{e}{m}\left[a {\vec B}+\left(\frac{\eta}{2}-\frac{\gamma^2}{\gamma^2-1}\chi \right)\frac{{\vec v}}{c}\times{\vec B}\right]\, .
    \label{restafreqmonopoles}
    \end{equation}

We recover the result obtained for the contribution of electric dipole moment, plus the contribution of magnetic monopoles. However, as discussed before, they do not contribute to the effective value of $a$.


\subsection{delta scenario}
\label{deltacaseexactsect}

Another interesting case might be chosing all parameters equal zero, except of
$(e, g, \delta)$.
In this case, the equations of motion read
\bea
m \frac{d p^\mu}{d \tau} &=&
\frac{e}{c} F^{\mu \nu}p_\nu
+\delta\, S^\mu (FSp) 
\label{deltacase1}\eea
and
\bea\label{eq_eomSdelta}
m \frac{d S^\mu}{d \tau}&=&
\frac{ge}{2c} F^{\mu \nu}S_\nu 
+\frac{e}{c\, m^2}\left(\frac{g}{2}-1-\delta\, S^2 \frac{c}{e}\right) p^\mu (FSp).
\eea
It's evident that the parameter $\delta$
directly alters the anomalous magnetic moment. 
 To show this in an explicit manner, an exact
solution is presented in this section. This can be calculated from the above equation by defining the new antisymmetric tensor
\begin{equation}
    \xi^{\mu\nu}=F^{\mu\nu}+\frac{\delta c}{e}\left(S^\mu S_\alpha F^{\alpha\nu}-S^\nu S_\alpha F^{\alpha\mu} \right)\, .
\end{equation}
Using this tensor, Eq.~\eqref{deltacase1} can be written in a simplified form as
\begin{eqnarray}
m \frac{d p^\mu}{d \tau} =
\frac{e}{c} \xi^{\mu \nu}p_\nu\, .
    \label{deltacase3}
\end{eqnarray}
Similarly, after a lenghty algebra exercise, it can be shown that Eq.~\eqref{eq_eomSdelta} can be written as
\bea\label{deltacase4}
m \frac{d S^\mu}{d \tau}&=&
\frac{{\hat g}e}{2c} \xi^{\mu \nu}S_\nu 
+\frac{e}{c\, m^2}\left(\frac{{\hat g}}{2}-1\right) p^\mu (\xi Sp)\, ,
\eea
where $\xi S p\equiv S_\alpha\xi^{\alpha\beta}p_\beta$, and 
\begin{equation}
    \hat g=\frac{g}{1+c\,\delta\, S^2/e}\, .
    \label{effectgdelta}
\end{equation}

We can see that Eqs.~\eqref{deltacase3} and \eqref{deltacase4} are completly equivalent to Eqs.~\eqref{eq_BMT1} and \eqref{eq_BMT2}, by changing $F^{\mu\nu}\rightarrow \xi^{\mu\nu}$, and $g\rightarrow\hat g$. Therefore, it is clear from 
Eq.~\eqref{effectgdelta} that spin modifies effectively the anomalous magnetic moment. This can be shown by following the procedure outlined in Sec.~\ref{ss_BMT}. For the delta case scenario,  using Eqs.~\eqref{deltacase3} and \eqref{deltacase4} in the case of no electric fields,
the momentum and spin oscillation frequencies vectors are respectively
\bea
{\vec \omega_{p\xi}}&=& \frac{e}{m c \gamma} \vec B_\xi, \label{freqdeltasce1}\\
\vec \omega_{\zeta\xi}&=& \frac{{\hat g}e}{2 m c \gamma} \vec B_\xi +  ({\hat g}-2)
\frac{e}{2 m c}\frac{\gamma}{\gamma +1} \frac{\vec v^2}{c^2} \vec B_\xi ,\label{freqdeltasce2}
\eea
where $\vec B_\xi$ is the magnetic part of tensor $\xi^{\mu\nu}$, being
\begin{eqnarray}
    \vec B_\xi=\vec B-\frac{c\,\delta}{e}\vec S\times\left(\vec S\times\vec B\right)\, .
    \end{eqnarray}
Both frequencies are parallel to ${\vec B}_\xi$, fulfilling 
$\vec v\cdot \vec B_\xi=0$. Thus, we can now define two ways to measure the anomalous magnetic moment. 

As a first option, following definition \eqref{eq_amu1}, 
we can  define it as the projection of the substraction of the above frequencies along the magnetic part of the tensor $\xi^{\mu\nu}$. In this case
\be\label{deltaanomalousmagdelta1}
a_{\mu\xi}= \frac{m c}{e\vec B_\xi^2 }
(\vec \omega_{\zeta\xi} - \vec \omega_{p\xi})\cdot \vec B_\xi= \frac{{\hat g}-2}{2}.
\ee
Under this definition, the usual anomalous magnetic moment differs from \eqref{deltaanomalousmagdelta1} by the coupling of the spin dynamics through the delta factor, implying that
\begin{eqnarray}
    a_\mu-a_{\mu\xi}=\frac{g}{2}\left( \frac{c\, \delta S^2/e}{1+c\, \delta S^2/e}\right)\, .
\end{eqnarray}

As a second case, the anomalous magnetic moment can be also defined as a projection on the magnetic field of the difference of the oscillation frequencies, as
\be\label{deltaanomalousmagdelta2}
a_{\mu\xi}= \frac{m c}{e\vec B^2 }
(\vec \omega_{\zeta\xi} - \vec \omega_{p\xi})\cdot \vec B= \frac{{\hat g}-2}{2}\left(1+\frac{c\, \delta\, S^2}{e}-\frac{c\, \delta}{e}\frac{(\vec S\cdot\vec B)^2}{\vec B^2} \right).
\ee
Both definitions \eqref{deltaanomalousmagdelta1} and \eqref{deltaanomalousmagdelta2} produce different results depending on the knowledge on the physical variables of the studied dynamics, for example, on the variables controlled in an experiment.  However, both results show that the $\delta$ parameter modified the anomalous magnetic moment due to the spin. Thus, any measured
difference with respect to \eqref{eq_amu1} could be provoked by a $\delta$ coupling.

The above results demonstrate that corrections to the anomalous magnetic moment arise, contingent upon the initial conditions set for the particle's spin and momentum. Also, the above problem will be  tackled through a perturbative analysis in Sec.~\ref{scenardelta}. 

\section{Polishing II,  perturbation theory}
\label{sec_perturbation}

Both, the number of free couplings and
the complexity of the coupled equations poses challenges in deriving exact results for tangible observables, such as $g-2$.
In this section, we employ an automatable perturbative approach to assess the phenomenological significance of the myriad potential couplings and corrections. We'll examine this within the context of a simplified yet substantive toy scenario.

\subsection{Unperturbed solution}

As simplest scenario for a perturbative approach we assume that all corrections vanish, including the anomalous magnetic moment, thus $g=2$,
as obtained from the tree level Dirac equation.
Since the unperturbed equations are sometimes  in the literature written in terms of a four velocity $u_0^\mu$ instead of a four momentum $P_0^\mu$ we define $P_0^\mu \equiv m u_0^\mu$ with $u_0^\mu u_\mu^0=1$. Both are equivalent for this particular case.
The uperturbed equations read
\bea
m \frac{d u_0^\mu}{d \tau} &=&
\frac{e}{c} F^{\mu \nu}u_{\nu,0}
\eea
and
\bea
m \frac{d S^\mu_0}{d \tau}&=&
\frac{e}{c} F^{\mu \nu}S_{\nu,0}
\eea
For  constant magnetic field and vanishing electric field
\bea
\vec B &=& \hat z B_z\\
\vec E &=&0,
\eea
the simplest solution is
\bea\label{eq_Pmu0}
P_0^\mu &=& \left(
\begin{array}{c}
E\\ 
P_{x,m} \sin \left( \tau e B_z /(m c)\right)\\
P_{x,m} \cos (\tau e B_z /(m c))\\
0
\end{array}
\right),
\eea
and
\bea\label{eq_S0}
S_0^\mu &=& \left(
\begin{array}{c}
\frac{P_{x,m} S_{x,s}}{E}\\
S_{x,c} \cos (\tau e B_z /(m c))+
S_{x,s} \sin (\tau e B_z /(m c))\\
-S_{x,c} \sin (\tau e B_z /(m c))+
S_{x,s} \cos (\tau e B_z /(m c))\\
\sqrt{S^2-S_{x,c}^2-\frac{m^2}{E^2} S_{x,s}^2}
\end{array}
\right),
\eea
where $E^2=m^2+P_{x,m}^2$.
The solution (\ref{eq_S0}) is easily translated into the local spin pseudo vector 
\be\label{eq_zeta0}
\vec \zeta_0=\left(
\begin{array}{c}
S_{x,c}\cos(\tau e B_z /(m c))
+\frac{m}{E}S_{x,s}\sin(\tau e B_z /(m c))\\
\frac{m}{E}S_{x,s}\cos(\tau e B_z /(m c))-S_{x,c}\sin(\tau e B_z /(m c))\\
\sqrt{S^2-S_{x,c}^2-\frac{m^2}{E^2}S_{x,s}^2}
\end{array}
\right).
\ee

\subsection{Perturbative corrections}

The full eight equations
(\ref{eq_eom2U}, \ref{eq_eom2S}) 
can be reduced to six equations
by virtue of the conserved mass and spin. The remaining dynamical variables 
are then $\vec P(\tau)$ and $\vec S(\tau)$, or equivalently the six variables
$\vec P(\tau)$ and $\vec \zeta(\tau)$.
Seeking small deviations from
the classical solutions (\ref{eq_Pmu0}, \ref{eq_zeta0}) we expand the general vectors as
\bea
\vec P&=& \vec P_0 + \epsilon  \vec P_1\\
\vec \zeta&=& \vec \zeta_0 + \epsilon  \vec \zeta_1,
\eea
where the the small modifications are given in terms of the functions ($\vec P_1(\tau)$ and $\vec \zeta_1(\tau)$)
.
Further, to keep track of the corrections with respect to the leading order classical result we equip each remaining correction prefactor with the expansion parameter $\epsilon$.
This expansion parameter will be set to one at the end of the calculation.
We find
\bea\label{eq_PertCoeff}
\alpha^* &\rightarrow &\epsilon \alpha^*, \\ \nonumber
\delta^* &\rightarrow &\epsilon \delta^*, \\ \nonumber
D^* &\rightarrow &\epsilon D^*, \\ \nonumber
\delta &\rightarrow &\epsilon \delta, \\ \nonumber
\beta^* &\rightarrow &\epsilon \beta^*, \\ \nonumber
A^* &\rightarrow &\epsilon A^*, \\ \nonumber
g &\rightarrow &2 \left(1 + \epsilon \tilde A\right), \\ \nonumber
\eea
where the usual contribution to the anomalous magnetic moment is defined as
\be\label{eq_gm2Def}
\tilde A= \frac{g-2}{2}.
\ee
Inserting the expansion (\ref{eq_PertCoeff}) into the six full equations
(\ref{eq_eom2U}, \ref{eq_eom2S}) 
the right hand side of these equations can be ordered as a power expansion in $\epsilon$. To leading order,
the equations are solved by (\ref{eq_Pmu0}, \ref{eq_zeta0}). 
To subleading order, these are
six coupled linear differential equations for $\vec P_1$ and $\vec \zeta_1$.
It is thus useful to define the six component object
\be\label{eq_DefQ}
Q^J(\tau)\equiv \left(\vec P_1,\, \vec \zeta_1\right),
\ee
where the capital letter index $J$ runs from one to six.
With this, we can write the equations to subleading order as
\be\label{eq_eomQ}
\frac{d Q^J}{d\tau}= M^J_{\; K} Q^{K}(\tau),
\ee
where $M^J_{\; K}$ is a $6 \times 6$ matrix, with coefficients given in terms of the unperturbed solution (\ref{eq_Pmu0}, \ref{eq_zeta0}) and the perturbation coefficients (\ref{eq_PertCoeff}).
This matrix is independent of the dynamical variable $Q^J$. Thus, the equations (\ref{eq_eomQ}) are truly linear in the dynamical variable $Q^J(\tau)$, which is beneficial for the computation.
For example,
the eigenvalues of the matrix contain then the information on possible modifications of the eigenfrequencies $\Omega_{J}$ of
the coupled spin-momentum system. We have derived the matrix $M_{\;K}^{J}$ in a computer algebra program, which is added as complementary material, but it is too large to type it in readable form since it fills many pages.
Thus, obtaining general analytic eigenvalues for $6 \times 6$ matrices remains elusive. Our recourse is to either numerical methods or to analytical evaluations for specific benchmark scenarios. In the following section, we present the analytic outcomes for each individual modification as specified in (\ref{eq_PertCoeff}).
In one of these sections we will present a (particularly simple) example for the matrix $M_{\;K}^{J}$.

\subsubsection{Scenario: $g-2$ or $\tilde A$}

The most prominent application of the BMT equations is the measurement of the anomalous magnetic moment $g-2$ (in addition to $e, \; m,\; S^2$).
This is achieved by a comparison between the 
spin precession frequency and the cyclotron frequency.
The cyclotron frequency is directly obtained from the eigenvalue
of the RHS of the $P^\mu$ equation.
The spin precession frequency is obtained as
eigenvalue of the pseudo-vector (\ref{eq_zeta}).
In the usual scenario the equations
for the vectors $\vec \zeta$ and $\vec v$ 
can be readily decoupled and one has
thus direct access to the eigenvalue of $\vec \zeta$.
In the light of the generalized equations (\ref{eq_DefQ}), it is however not clear how to decouple the equations.
However, we have have straight forward perturbative access to the eigenvalues 
of $\vec \zeta_1$ and $\vec P_1$.

To study the usual $g-2$ correction 
in the light of the generic scenario with
a $6\times 6$ matrix (\ref{eq_eomQ})
we set all $\sim \epsilon$ modifications in (\ref{eq_PertCoeff}) to zero, except of $e$ and $g$. 
Next, we compute the eigenvalues of the $6\times 6$ matrix present on the right-hand side. To leading order, and when averaged over a period of the unperturbed solution, the matrix has four non-zero eigenvalues:
\be\label{eq_EVOm}
\Omega_{J}|_g=
\left\{
\pm i \frac{e}{m} B_z, \pm i \frac{e}{m} B_z \left( 1+\frac{E(g-2)}{2m}
\right)
\right\}.
\ee
The first two $\Omega_{1,2}$
are the frequency of the momentum $\vec P(\tau)$
the last two $\Omega_{3,4}$ 
are the frequency of $\vec \zeta(\tau)$.
Thus, by subtracting the measurable $\Omega_{1}$ and $\Omega_{3}$, dividing by $ie B_z/m$ and multiplying a kinematic factor of $m/E$ one can extract the well-known $(g-2)/2$ as
\be\label{eq_gm2Omega}
a|_{obs}\equiv \frac{\Omega_3-\Omega_1}{ie B_z}\cdot \frac{m^2}{E},
\ee
where the kinematic factor $m/E$ arises when one goes from proper time to laboratory time $\tau \rightarrow t$.
Inserting the eigenvalues (\ref{eq_EVOm})
into the definition (\ref{eq_gm2Omega})
we confirm readily that the perturbative analysis 
\be\label{eq_gm2pertg}
a_{obs}|_{g}=\frac{g-2}{2},
\ee
in agreement with the non-perturbative result (\ref{eq_gm2usual})
This outcome serves as a robust consistency check for our approach, centered on the eigenvalues of the matrix $M^J_K$. However, it's imperative to recognize that this strategy, while illuminating, is a marked oversimplification when juxtaposed with authentic experimental situations. Despite this, its redeeming quality lies in its applicability across all the subtle alterations presented in (\ref{eq_PertCoeff}).

\subsubsection{Scenario: $\alpha^*$, Magnetic monopoles}

To study the $\sim \alpha^*$ correction we set all $\sim \epsilon$ modifications in (\ref{eq_PertCoeff}) to zero, except of $\alpha^*$. The resulting averaged matrix
is
\be
\langle M_{\;K}^J\rangle=\epsilon\left(\begin{array}{cccccc}
    0 & \frac{B_z e}{m} & 0 & 0 & 0 & 0  \\
    -\frac{B_z e}{m} & 0 & 0 & 0 & 0 & 0  \\
    0 & 0 & 0 & 0 & 0 & 0  \\
    M_4^1 & M_4^2 & 0 & 0 & \frac{B_z e}{m} & 0  \\
    -M_4^2 & M_4^1 & 0 & -\frac{B_z e}{m} & 0 & 0  \\
    0 & 0 & 0 & 0 & 0 & 0 
\end{array}\right)
\ee
where
\bea
M_4^1&=&\frac{B_z (m^2+E_0 m-E_0^2)\sqrt{S^2 -S_{x,c}^2-m^2 S_(x,s)^2/E_0^2}\alpha^*}{m^2 (E_0+m)},\\ \nonumber
M_4^2&=&-\frac{B_z e(E_0^2-m^2)^{3/2} S_{x,s}}{E_0^2m\sqrt{E_0+m}}.
\eea
There are 
only four non-vanishing averaged
eigenvalues of the matrix to leading order
\be
\Omega_{J}|_{\alpha^*}=
\left\{
\pm i \alpha B_z,\pm i \alpha B_z
\right\}.
\ee
The parameter $\alpha^*$ dropped out of these eigenvalues. 
Thus, evaluating the observable (\ref{eq_gm2Omega})
shows that the value of $\alpha^*$ has no impact on the measurement of the anomalous magnetic moment
\be\label{eq_gm2alphas}
a_{obs}|_{\alpha^*}= \frac{\Omega_3-\Omega_1}{ie B_z}\cdot \frac{m^2}{E}=0.
\ee
This is in agreement with the analytical result for monopoles found in Eq.~(\ref{restafreqmonopoles}), in which is shown that magnetic monopoles do not modify the effective value of $a$.

Now, that we have checked that the perturbative analysis gave the correct results for the scenarios 
(\ref{eq_gm2pertg}) and (\ref{eq_gm2alphas}), we proceed to explore the remaining modifications.

\subsubsection{Scenario: $\beta^*$}

To study the $\sim \beta^*$ correction we set all modifications in (\ref{eq_PertCoeff}) to zero, except of $\beta^*$.
Interestingly there are five non-vanishing averaged eigenvalues of the matrix  to leading order
\bea
\Omega_{J}&=&
\left\{
\pm \frac{ B_z}{m^2}\left(i em +\sqrt{E^2(S^2-S_{x,c}^2)-m^2 S_{x,s}^2}\beta^*\right),\right.\\ \nonumber &&\left.
\pm \frac{ B_z}{m^3}
\left(i em^2 - E (E-m) \sqrt{S^2-S_{x,c}^2-\frac{m^2}{E^2}S_{x,s}^2}\beta^* \right),
-\frac{B_z}{m^2}\sqrt{E^2(S^2-S_{x,c}^2)-m^2 S_{x,s}^2}\beta^*
\right \}
\eea
where $\Omega_5$ vanishes as $\beta^* \rightarrow 0$, while the other four eigenvalues simply go to $\pm i e B_z/m$. Now, 
applying the usual formula of differences one finds
\be
a_{obs}|_{\beta^*}
= \frac{\Omega_3-\Omega_1}{ie B_z}\cdot \frac{m^2}{E}=
-i \sqrt{E^2(S^2-S_{x,c}^2)-m^2 S_{x,s}^2}\frac{\beta^*}{e m}.
\ee
This indicates that small $\beta^*$ corrections can modify the $g-2$ observable. However, since this correction is imaginary, it means that the correction does not correspond to an oscillatory behaviour.

\subsubsection{Scenario: $\delta$}
\label{scenardelta}

To study the $\sim \delta$ correction we set all modifications in (\ref{eq_PertCoeff}) to zero, except of $\delta$.
There are now five non-vanishing eigenvalues of the matrix 
\bea
\Omega_{J}&=&
\left\{
\pm \frac{i B_z}{2}\left(
\frac{2 e}{m}-(S_{x,c}^2+S_{x,s}^2)\delta
\right),\right. \\ \nonumber&&
\left.
\pm \frac{i B_z }{2 E^3 m}
\left(
2 E^3 e +(E-m)(
E^3(S^2-S_{x,c}(S_{x,c}+i4S_{x,s}))
+m^3 S_{x,s}^2- E^2 m (
S_{x,c}^2-i S_{x,c} S_{x,s}+S_{x,s}^2
)
)\delta
\right),\right. \\ \nonumber&&
\left.
\frac{B_z}{E}(E-m)S_{x,c} S_{x,s}\delta
\right\},
\eea
of which only four are also present at leading order. This means that 
$\Delta \Omega$ has to be formed with the appropriate two of these four. To make the somewhat lengthy result more readable we evaluate it in the zero momentum limit where $E=m$
\bea
a_{obs}|_{\delta, E=m}
&=& \frac{\Omega_3-\Omega_1}{ie B_z}\cdot \frac{m^2}{E}\\ \nonumber
&=&\delta
\frac{m(S_{x,c}^2+ S_{x,s}^2) }{2 e}.
\eea
One finds that $\delta$
indeed mimics an oscillatory 
motion of the type one would obtain
for $g-2$. This in agreement with the 
result  of Sec.~\ref{deltacaseexactsect}.

\subsubsection{Scenario: $\delta^*$}

To study the $\sim \delta^*$ correction we set all  modifications in (\ref{eq_PertCoeff}) to zero, except of $\delta^*$.
There are now six non-vanishing eigenvalues of the matrix are to leading order, resulting in a large algebraic expression. To keep it more readable we give the result in the low energy limit
\be
E_0\approx m + E_{k},
\ee
where the kinetic energy is much smaller than the invariant mass $E_k\ll m$.
The eigenvalues are then
\bea
\Omega_{J}|_{\delta^*, E_k}&=&
\left\{
\pm \frac{i B_z e}{m },\right.
\\ \nonumber &&
\left.
\pm \frac{B_z}{4m^2}
\left(4 i e m+ \sqrt{2}(5 S_{x,s}-iS_{x,c}) \delta^* \sqrt{m^3(-|S^2|+S_{x,s}^2+S_{x,c}^2)E_k^2}\right)\right. \\ \nonumber &&
\left.
\pm \sqrt{2}B_z S_{x,s} \delta^*
\sqrt{-\frac{-|S^2|+S_{x,c}^2+S_{x,s}^2}{m}E_k}
\right\},
\eea
of which only the first four are also present at $\delta^* \rightarrow 0 $. 
Calculating the $g-2$ observable, one finds in this limit
\bea
a_{obs}|_{\delta^*, E_k}
&=& \frac{\Omega_3-\Omega_1}{ie B_z}\cdot \frac{m^2}{E}\\ \nonumber
&=&
-\frac{(S_{x,c}+ 5 i S_{x,s})\delta^* \sqrt{-m (S^2+S_{x,c^2}^2+S_{x,s}^2)E_k}}{2 \sqrt{2}e},
\eea
where $S^2=-|S|^2$.
This shows, that also $\delta^*$ corrections can
mimic an oscillatory 
motion of the type one would expect
for $g-2$.

\subsubsection{Scenario: $A^*$}

To study the $\sim A^*$ correction we set all  modifications in (\ref{eq_PertCoeff}) to zero, except of $A^*$.
For this scenario we find
four non vanishing, but degenerate
eigenvalues of the matrix $M^I_J$
\bea
\Omega_{J}|_{A^*}&=&
\left \{ \pm \frac{i B_z e}{m}, \pm \frac{i B_z e}{m}\right\}.
\eea
Thus, there are no contributions
to the $g-2$ observable from $A^*$ corrections
\bea
a_{obs}|_{A^*}
&=&0.
\eea
\subsubsection{Scenario: $D^*$}

Finally, to study the $\sim D^*$ correction we set all  modifications in (\ref{eq_PertCoeff}) to zero, except of $D^*$.
For this scenario we find
five non vanishing
eigenvalues of the matrix $M^I_J$
\bea
\Omega_{J}|_{A^*}&=&
\left \{ \pm \frac{i B_z e}{m}, \right.
\\ \nonumber &&
\left.
\pm B_z
\left( 
\frac{ie}{m}- D^*\sqrt{E^2(|S|^2-S_{x,c}^2)-m^2 S_{x,s}^2}
\right),- 2 B_z D^*\sqrt{E^2(|S|^2-S_{x,c}^2)-m^2 S_{x,s}^2}
\right\}.
\eea
The fifth eigenvalue vanishes in the limit $D^* \rightarrow 0$.
The $g-2$ observable is then found
to be
\bea
a_{obs}|_{D^*}
&=&  \frac{\Omega_3-\Omega_1}{ie B_z}\cdot \frac{m^2}{E}=\frac{i D^* m^2}{E e}\sqrt{E^2(|S|^2-S_{x,c}^2)-m^2 S_{x,s}^2}.
\eea
This corresponds to
a non-oscillatory contribution to
the observable of the anomalous magnetic moment.

\section{Discussion}

\subsection{Dependence on initial conditions}

Several results presented previously are influenced by the initial conditions of the spin, momentum, and energy. Such dependencies are anticipated since similar behaviors are observed in the standard BMT theory. The most general result for the subtraction of the proper characteristic frequencies, from BMT theory given by Eqs~\eqref{eq_BMT1} and \eqref{eq_BMT2}, is \cite{Jackson,Muong-2:2021ojo}
\bea
\vec \omega_\zeta-\vec \omega_p&=& 
\frac{e}{mc}\left[
\left(\frac{g}{2}-1\right) \vec B
 -\left(\frac{g}{2}-\frac{\gamma^2}{\gamma^2-1}\right)\vec v\times\vec E-\vec v\frac{\gamma}{\gamma+1}\left(\frac{g}{2}-1\right) \vec v\cdot \vec B
\right].
\label{BMTgentotal}
\eea
Consequently, the definition for measuring the anomalous magnetic moment, as given by \eqref{eq_amu1}, is contingent upon an apt selection of the initial condition. An illustrative example of this can be found in Ref.~\cite{Muong-2:2021ojo}.  The experimental setup discussed in \cite{Muong-2:2021ojo} 
is meticulously designed  to minimize the impact of the additional terms present in Eq.~\eqref{BMTgentotal}. In principle, it is thus useful to choose a configuration such as $\vec v\cdot\vec B=0$, and $\gamma\approx \sqrt{g/(g-2)}\approx 29.3$. In this way, the previous result \eqref{BMTgentotal} simplifies to $\vec \omega_\zeta-\vec \omega_p= ({e}/{mc})({g}/{2}-1) \vec B$, which allows to deduce the anomalous magnetic moment using Eq.~\eqref{eq_amu1}.
However, experimentally, it is not possible to realize exactly the orthogonality between $\vec v$ and $\vec B$. Thus, neither the second nor the third term of \eqref{BMTgentotal}  vanish exactly. The experimental setup ensures that these terms are small, considering always their small impact  in the analysis and the uncertainty estimation.

While our findings may initially appear counter intuitive, given that fundamental constants should not be contingent upon initial conditions, there is no inherent contradiction. This is because the formulas employed to determine these constants are predicated on a particular theoretical foundation, namely, the equations of motion. As our proposed equations of motion diverge from traditional models, it's logical that identical cancellations, which might be anticipated in other contexts, don't necessarily transpire when applying a relation tailored for a distinct equation of motion.

\subsection{Imaginary eigenvalues of eigen frequencies}

Another noteworthy feature of our results is the presence of imaginary parts in some eigenvalues. This signifies a growing deviation from the standard oscillatory motion. Such findings are anticipated. In all interactions with the dual field strength tensor, the roles of the electric and magnetic fields are swapped. Within these interactions, the presupposed constant uniform magnetic field behaves analogously to a constant uniform electric field, which aligns with the concept of unbounded acceleration.
Evidently, the described field configuration is a marked oversimplification when compared to genuine experimental settings. In more realistic situations, particles are restricted in every direction. As a result, confined oscillatory motion, for instance, in the $\hat z$ direction, should prevail, especially given that all corrections given by (\ref{eq_PertCoeff}) are anticipated to be negligible.

\subsection{Are the corrections predicted by the Standard Model?}

If one thinks about fundamental particles, our novel corrections are exactly zero at tree level from the perspective the Standard Model.
If they are induced at loop level they will be strongly suppressed and most likely not measurable.

However, they are perfectly viable for
\begin{itemize}
    \item Composite particles:\\
    Bound states of multiple particles can have a very rich internal structure, mutipole- and intertial moments, which would be absent for the fundamental particles present in the Standard Model. The presence of multipole moments of composite particles like deuteron is well known and well studied in scattering and decay processes~\cite{Bernstein:1979zza,Knupfer:1980un,Arenhovel:2002sg,Arenhovel:1995be}. 
    The relations (\ref{eq_eom2U}) offer a structured way to implement multipole interaction with electromagnetic fields in the context of the BMT equations.
    Thus, for each composite particle species, the coefficients in the equations (\ref{eq_eom2U}) should be  constrained or determined by separate experiments.
    Clearly, the multipole moments of composite particles, which are held together by the strong interaction are far smaller than those of more loosely bound particles. Nevertheless, it would be interesting to investigate to which extent hadronic storing experiments, like COSY or JEDI~\cite{Chiladze:2005qw,JEDI:2022hxa} could test and constrain the new coefficients.
    \item Beyond the Standard Model (bSM):\\
    Any extension of the SM can, through an interaction with SM particles, generate some non vanishing coefficients in the equations (\ref{eq_eom2U}).
    The possibilities for bSM extensions are countless, 
    but the existence of bSM physics is an experimental fact, as it is well knwon in the context of neutrino physics. While the neutrinos of the SM are massless by construction, real neutrinos do actually have mass. There are multiple bSM extension trying to explain this experimental fact and depending on the extension, some coefficients of the equations (\ref{eq_eom2U}) are non-vanishing. For example,
    it is investigated whether the neutrino carries anomalous  moments.
    Clearly, measuring the anomalous magnetic moments of neutrinos is experimentally not possible at the time. Other moments,
    such as for example $\alpha^*$ which can be associated to magnetic monopole features are probably more promising in the context of neutrinos~\cite{Carrigan:1971mx,Okulov,Steyaert:1987zh,Capelas,lochak,Jeong:2021kvw}. Now, the equations (\ref{eq_eom2U}) make it possible to test this type of hypothesis systematically. 
     In particular, they allow to predict how much a long baseline neutrino beam~\cite{LBNE:2013dhi}
    would be deflected when an electromagnetic field is applied. The resulting eventually measurable deflection would depend on the parameter $\alpha^*$.  
    This, in turn, would allow to explore or constrain this parameter space.
    The same type of result and argument would also arise (in a more complicated form) for different parameters such as $\{\beta^*, \;\delta,\, \delta^*,\, \tilde \alpha^*,\, \tilde \beta^*, \dots  \}$.
\end{itemize}

\section{Conclusions}

In this study, we revisited the BMT equations pertaining to the spin and momentum of a particle. Our approach encompassed all possible Lorentz invariant corrections that are first order in the electromagnetic field strength tensor. 
We then studied the dynamics of these equations for the special case of vanishing gradients of external fields.
Our findings identified six couplings that extend beyond the conventional interpretation of these equations.

We then explored exact and approximated results for these equations, focusing on the orbiting and spin precession
in a constant magnetic field.
The comparison between the two corresponding frequencies is a standard candle for the anomalous magnetic moment. The results of our perturbative analysis on this observable is summarized in table \ref{tab:summary}.
\begin{table}[h!]
    \centering
    \begin{tabular}{|c|c|c|c|c|c|c|c|}
        \hline
Correction      & $g-2$ & $\alpha^*$ & $\beta^*$ & $\delta$ & $\delta^*$ & $A^*$& $D^*$\\
         \hline
Contribution   & y & n& y$^*$ & y & y & n& y$^*$ \\
         \hline
    \end{tabular}
    \caption{Summary on whether the corrections could contribute to the observable defined in (\ref{eq_gm2Omega}.) The letter ``y'' indicates yes, the letter ``n'' indicates no, and the letter ``y$^*$'' indicates a  non-oscillatory contribution.}
    \label{tab:summary}
\end{table}

From a phenomenological perspective, our main result is:
``Whenever some of the new couplings in (\ref{eq_eom2U}) and (\ref{eq_eom2S}) do exist in nature, they alter the dynamics 
and thus the interpretation of observed experimental quantities,
as it is summarized in table \ref{tab:summary}.''
For future projects it would be interesting 
to explore how our parameters arise from an effective quantum field theory perspective~\cite{Brivio:2017vri} of fundamental particles such as muons~\cite{Muong-2:2021ojo,Muong-2:2023ojo} or from a perspective of composite nature of the particle under consideration.

\begin{acknowledgments}
The authors express their gratitude to Sergio Hojman and Anton Rebhan for their insightful comments.
B.K. was supported by the grants P 31702-N27 and P 33279-N.
\end{acknowledgments}

\end{document}